\documentclass[conference]{IEEEtran}
\IEEEoverridecommandlockouts
\usepackage{cite}
\usepackage{amsmath,amssymb,amsfonts}
\usepackage{algorithmic}
\usepackage{graphicx}
\usepackage{comment}
\usepackage{textcomp}
\usepackage{xcolor}
\def\BibTeX{{\rm B\kern-.05em{\sc i\kern-.025em b}\kern-.08em
    T\kern-.1667em\lower.7ex\hbox{E}\kern-.125emX}}
\begin{document}

\title{The 10 Research Topics in the Internet of Things}

\author{\IEEEauthorblockN{Wei Emma Zhang$^1$, Quan Z. Sheng$^2$, Adnan Mahmood$^2$, Dai Hoang Tran$^2$, Munazza Zaib$^2$, \\Salma Abdalla Hamad$^2$, Abdulwahab Aljubairy$^2$, Ahoud Abdulrahmn F. Alhazmi$^2$, Subhash Sagar$^2$, and Congbo Ma$^1$}
\IEEEauthorblockA{$^1$School of Computer Science, The University of Adelaide, SA 5005, Australia \\
Email: wei.e.zhang@adelaide.edu.au\\
$^2$Department of Computing, Macquarie University, NSW 2109, Australia\\
Email: michael.sheng@mq.edu.au
}
}


\maketitle

\begin{abstract}
Since the term first coined in 1999 by Kevin Ashton, the Internet of
Things (IoT) has gained significant momentum as a technology to connect
physical objects to the Internet and to facilitate machine-to-human and machine-to-machine communications. 
Over the past two decades, IoT has been an active area of research and development endeavors by many technical and commercial communities. Yet, IoT technology is still not mature and many issues need to be addressed. 
In this paper, we identify 10 key research topics and discuss the research problems and opportunities within these topics.
\end{abstract}

\begin{IEEEkeywords}
Internet of Things, Energy Harvesting, Recommendation, Search, Summarization, Conversational IoT, IoT Service Discovery \end{IEEEkeywords}


\section{Introduction}
The vision of a connected and smart world can be traced back to 1920s, as explained by Nikola Tesla in 1926: 
``{\em When wireless is perfectly applied the whole earth will be converted into a huge brain, which in fact it is, all things being particles of a real and rhythmic whole. We shall be able to communicate with one another instantly, irrespective of distance. $\dots$ A man will be able to carry one in his vest pocket}''~\cite{Tesla}. However, the term ``Internet of Things'' (IoT) was only first coined in 1999 by MIT's Kevin Ashton when he promoted the radio frequency identification (RFID) technology.
Since then, IoT has received significant momentum as a promising technology to turn each physical object (i.e., a thing) into a node on the Internet and to facilitate machine-to-human and machine-to-machine communication with the physical world~\cite{TranSBYZD19,Welbourne-IC09,Sheng-book-WoT}. 
By connecting and integrating both digital and physical entities, IoT enables a whole new class of exciting applications and services such as smart cities, smart homes, Industry 4.0, and Society 5.0\footnote{https://www8.cao.go.jp/cstp/english/society5\_0/index.html}.

Over the past two decades, particularly the last 10 years, IoT has been a thriving area of research and development efforts, with a quickly rising body of produced research work. According to Microsoft Academic\footnote{https://academic.microsoft.com/topic/81860439.}, there were only 26 publications on IoT in 2000 and 160 publications in 2009. The publication number,  however, dramatically increased over the past few years, e.g., 10,926 in 2016, 15,765 in 2017, 21,906 in 2018, and 26,885 in 2019.  Despite of these exciting activities, IoT techniques still remain immature and many technical hurdles need to be overcome. 

The aim of this paper is to identify several important IoT research topics and areas, ranging from energy harvesting, data analytics, search, recommendation, security, privacy and trust in IoT, as well as the topics that arise in the adoption of new computing paradigms such as social computing, service computing, edge computing, and artificial intelligence (e.g., conversational AI and text/video summarization).

\begin{figure}[!t]
\begin{center}
\includegraphics[width=0.95\linewidth]{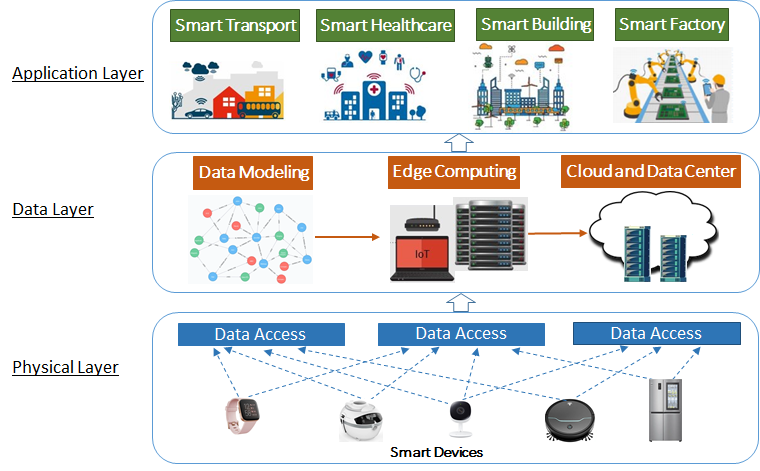}
\end{center}
\caption{ A generic architecture of IoT systems. The physical layer collects data to be processed and modeled in the data layer, which then delivers the data to the application layer.}
\label{fig:iot}
\end{figure}

\section{Research Directions in IoT}

Over the last two decades, there are a broad spectrum of activities around IoT research and development. In this paper, we highlight 10 topic areas that span 
across the three main layers of IoT architecture as  illustrated in Figure \ref{fig:iot}.
We believe these topics represent the most important research efforts from the community.
It should be noted that this is not the complete list of the important research topics. Many other topics such as standard development and regulatory implications are outside the scope of our discussions.

\subsection{Energy Harvesting}

The rapid evolution in the promising paradigm of IoT has resulted in a massive distributed network of intelligent objects possessing a highly varying compute, storage, and networking capabilities. These networked objects interact with one another primarily in a bid to exchange a diverse range of information having a direct influence for enhancing the quality of our daily lives by ensuring seamless access to smart services anywhere at anytime. However, a number of IoT sensors and embedded IoT devices have a limited lifespan since they are powered by batteries and, therefore, requires replacement periodically (e.g., every few years) making this an inefficient, laborious, and costly process. Smart energy management, especially, energy harvesting (also referred to as energy scavenging), is indispensable for ensuring energy efficiency in IoT objects \cite{Zeadally-EH}.

Energy harvesting is a mechanism for transforming readily available energy from natural or artificial resources into usable electrical energy. It comprises four salient phases, i.e., selection of an optimum and abundantly available energy resource, its transformation, storage, and consumption. Some of the energy sources that could be harvested for IoT include, but are not limited to, thermal energy, light energy, RF energy, electromagnetic energy,  chemical energy, and mechanical energy. For transformation purposes, corresponding harvesters or transducers are employed to identify and transform energy. In the case of storage, rechargeable batteries and super capacitors are exploited to store the energy. Finally, the harvested energy is consumed by appropriate IoT devices for their corresponding applications \cite{Sah-EH}. 

However, several underlying challenges still hinder the realization of an efficient IoT harvesting system. For instance, the harvesting circuitry has a considerable impact on the hardware of an IoT object since the conventional IoT objects' designs are unable to handle the heavy fluctuations in an object's circuitry, primarily owing to the fact that the harvested energy delivered to an IoT object is predominantly reliant on the availability of energy within the environment and which occasionally could be either inferior or even superior to the power requirements of an object's circuitry. Similarly, intelligent software for IoT harvesting systems should be designed by the software developers which are capable of handling the energy's unavailability for a shorter duration of time to allow any task to resume and not restart from where it was left, thereby mitigating the data loss. Finally, both rechargeable batteries and super capacitors have some inherent limiting factors, and therefore, a highly efficacious, economical, miniaturized, and long-lived battery is still a challenge for the researchers.


\subsection{Data-driven IoT}

IoT offers the capability to connect and integrate both digital and physical entities. 
A fundamental challenge centers around managing IoT data especially when things are the majority of data producers and consumers. Given the intrinsic features of IoT data, topics such as storage, real-time data stream analytic, and event processing are all critical. 
Before diving into these four topics, we would first summarize these features. 

Data generation in IoT has four main characteristics: i) \textit{Velocity}---things produce data in different speed levels and some sensors can scan at a rate up to 1,000,000 sensing elements per second\footnote{https://www.tekscan.com/support/faqs/what-are-sensors-sampling-rates}; ii) \textit{Scalability}---IoT data are expected to be at an extremely large scale due to the ability of IoT sensors to continuously generate data together with the foreseeable excessively large number of things; iii) \textit{Dynamics}---mobility is one characteristic of IoT things, leading to data generated in different locations under different environments at different times; and iv) \textit{Heterogeneity}---many kinds of things have been and could be connected to the Internet and the data generated could be in different formats using different vocabularies.

The quality of the generated data usually faces some special challenges. Data could cause uncertainty and inconsistency as sensors and RFID tags would produce inaccurate readings and redundant readings, or even miss readings.  Moreover, the data produced by assorted things can be interpreted in different ways, bringing challenges for proper interpretation of the produced data to different data consumers~\cite{Sheng-Computer08}. 

The nature of data produced by IoT calls for revisit of data storage techniques. Traditional datababase management systems (DBMSs) could be adopted for storing IoT data, but need to address the high processing and querying frequency. The   development of large-scale, distributed storage systems is also raised to meet the exceptional demands of data storage in IoT and three factors need to be considered: consistency, availability and partition tolerance \cite{monet/ChenML14}. The storage issue in resource-constrained IoT scenarios also plays an important role due to the mobility and scalability of IoT data. Antelope\footnote{https://github.com/contiki-os/contiki/blob/master/apps/antelope} is the first DBMS for resource-constrained sensor devices, which enables a class of sensor network systems where every sensor holds a database. 

Linked Data\footnote{http://linkeddata.org/} is a method for publishing structured data and interlink such data to make it more useful. It builds upon standard Web technologies such as HTTP, RDF and URIs and extends these technologies to share information. Data from different sources can be connected and queried in the form of Linked Data. The concept of Linked Stream Data applies the Linked Data principles to streaming data, so that data streams can be published as part of the Web of Linked Data. SPARQL is a query language to query RDF database and has been extended to support streaming data querying \cite{semweb/ZhangDCC12,semweb/CalbimonteCG10,www/AnicicFRS11}. Techniques in Linked Data perfectly fit the requirement of processing and querying IoT data at Web scale. 

In IoT, complex event processing techniques lay part of the foundation of supporting computers to sense and react to events in the physical world.  
The focus of the complex event processing (CEP) model is on detecting occurrences of particular patterns of low level events indicating some higher-level events, which are better understood by computers and humans.
A semantic CEP system improves event processing quality by using event metadata in combination with ontologies and rules (i.e., knowledge bases). A knowledge base can be used to provide background knowledge about the events and other non-event resources \cite{debs/TeymourianRP12}. 
With huge amount of external domain knowledge available, information completeness and semantic matching are two research topics under semantic CEP.  The process of reducing information incompleteness is called event enrichment which considers the challenges of enrichment source determination, information retrieval from the identified sources, and information fusion \cite{debs/HasanOC13}. Semantic matching includes semantic selection which is evaluating pattern constraints based on the semantic equivalence of attribute meanings captured by the event ontology, and inexact selection which is selecting events while allowing a limited number of mismatches to detect relevant patterns \cite{debs/ZhouSP11}. 

Although we see prompt development of IoT techniques, many challenges and issues 
are 
worth to further explore from both academia and industry.  Data quality and uncertainty remains a challenging problem due to the increasing data volume and heterogeneity. As in an IoT environment, the physical space and the virtual (data) space co-exist and interact. Data transferring, synchronizing and processing in the co-space demand novel techniques.  Given various formats of external knowledge, semantic enrichment will not only deal with structured data, but also unstructured data at the same time. How to enrich semantic in IoT via hybrid formatted external information is also an interesting topic to explore. In contrast to utilizing external knowledge, discovering knowledge from the IoT data could contribute to the understanding of the IoT applications. Data mining techniques could be exploited for this purpose but in a distributed manner. 

\subsection{IoT Search}

Searching and finding relevant objects from billions of things is one of the major challenges in the 
IoT era because 
the supporting technologies for searching things in IoT are very different from those used in searching Web documents due to 
tightly bounded contextual information (e.g., location) and no easily indexable properties of IoT objects. In addition, the state information of things is dynamic and rapidly changing. 

By reusing techniques of the World Wide Web, the information and services of IoT objects can be provided on the Web \cite{icess/DuquennoyGV09}. This triggers the research of Web of Things (WoT) search engines (WoTSE), which is applying Web technologies to the Internet of Things to access information and services of physical objects. In WoT, each physical object possesses a digital counterpart that is commonly referred to as 
``Digital Twin''.
These digital twins are built according to Representational State Transfer (REST) architecture and accessed with HTTP protocol via RestFul API. 

Research related to WoTSE begins from early 2000s and enjoys steady expansion ever since. 
It branches into different directions including object search, sensor search and functionality search \cite{csur/TranSBY17}. 
In early projects, WoTSE are commonly used to locate physical objects, which are tagged with passive RFID tags or sensor nodes. 
Dyser \cite{iot/OstermaierRMFK10} is one of the works that search physical entities based on their real-time states derived from their sensor readings. 
The work of 
CASSARAM \cite{mdm/PereraZCCG13} demonstrates the research effort on sensor search. It uses WoTSE for retrieving sensors based on their static meta-data and contexts, such as cost and reliability. 
Each form of WoTSE has its own characteristics, but all could follow the unified search architecture provided by our work in \cite{csur/TranSBY17}.

\begin{figure}[!t]
\begin{center}
\includegraphics[width=0.75\linewidth]{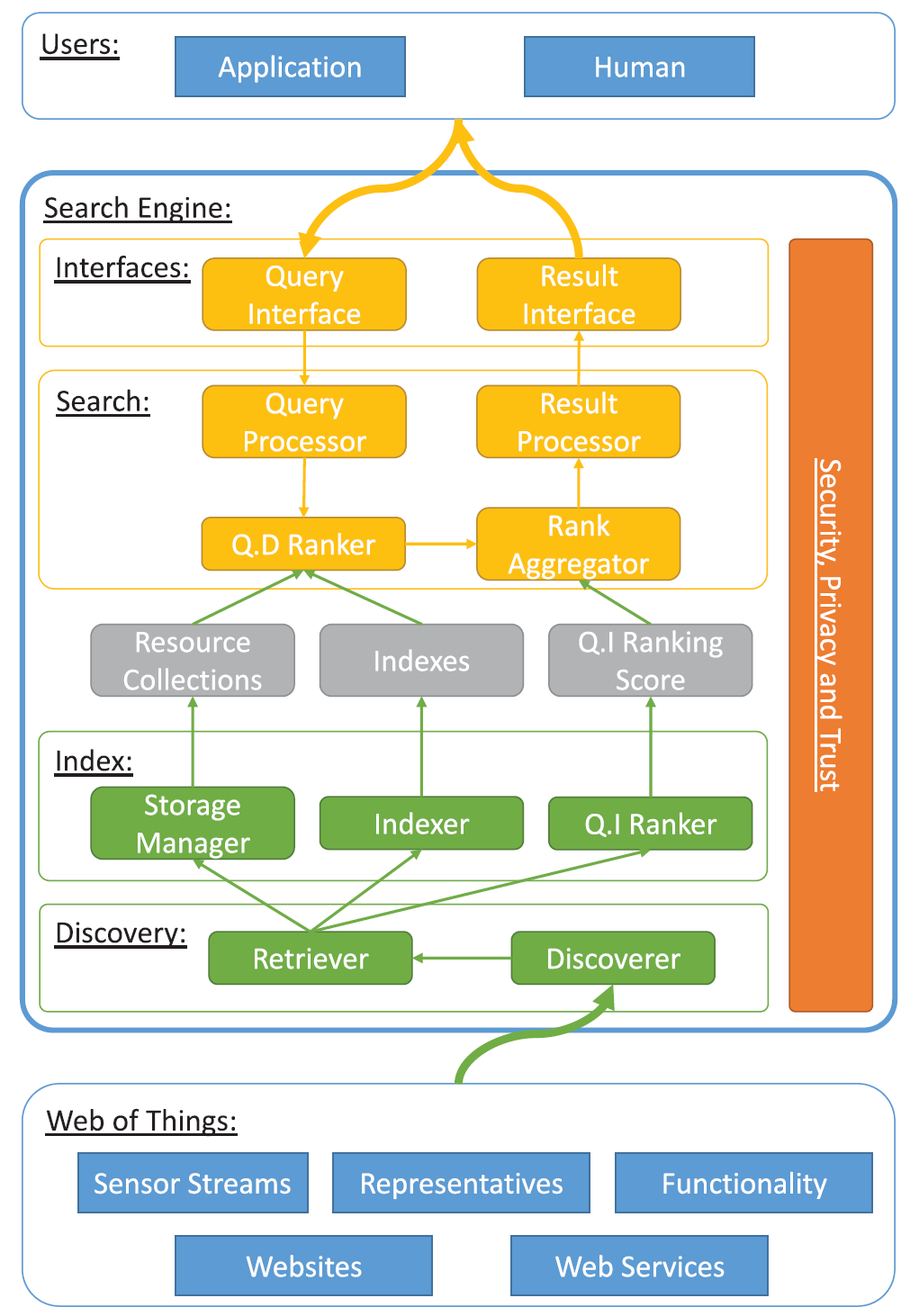}
\end{center}
\caption{Modular Architecture for Web of Things Search Engines \cite{csur/TranSBY17}.}
\label{fig:wotse}
\end{figure}

The modules in our architecture are organized into layers as illustrated in Figure \ref{fig:wotse}. Two lower layers handle discovery activities while the two 
upper layers handle search activities. 
Storage modules for resource collections and indexes link two set of layers. The whole system is protected by security, privacy, and trust assessment measures, which are grouped into a vertical layer. 
To be more specific, the Discovery Layer serves as interfaces to the Web resources including sensor streams, representations, functionalities, websites and Web services.
The Index Layer stores and indexes resources with its Collection Manager and Indexer modules.  This layer also ranks the resources. %
The Search Layer carries out the query resolution process. The Query Processor module transforms
raw user queries into the form processable by the system. The Query Dependent (Q.D) Ranker
scores discovered query resources with respect to the user query and utilizes the recorded links
between resources to find their corresponding result resources. The Ranking Aggregator module is responsible for combining different Q.D and Q.I ranking results into a final score for each resource. 
Finally, the Result Processor extracts and aggregates the information from matching  resources
and produces search results.
The User Interface (UI) layer interfaces WoTSE with users. It provides Query Interface and Result Interface
to receive queries and return search results, respectively. 

The modular architecture provides a reference framework assessing the diverse implementation
of the existing WoTSEs. It assesses the support that each module receives from the existing works and how it is commonly implemented.  
The goal of WoTSE is building a search engine that could find anything available on the Web of Things.  To achieve this goal, several challenges demand effective and efficient solutions. Crawling and indexing wide scale IoT data for the purpose of search are intrinsically 
problematic due to the dynamic and heterogeneous nature of IoT data. How to identify useful Web resources that are related to the things is also challenging as they could be in various format with different interpretation vocabularies as we discussed in the last section.    

\subsection{Security, Privacy, and Trust in IoT} 
The risk on data security and privacy exponentially increases with an unprecedented growth in the deployment of the smart IoT objects. One of the distinct challenges in the IoT infrastructures is the limited computation power and minimal resources of most of the IoT devices \cite{NIST8114}. These limited resources preclude the state-of-the-art cryptographic techniques that are indispensable for securing IoT devices, thereby making them vulnerable to a diverse range of security attacks \cite {Sain2017}, such as the denial of service attacks and privacy attacks such as data exfiltration or leakage attacks.

Recently, there are numerous research proposals in the literature delineating on IoT security and privacy services such as \cite{Touati2014, Ni2016, Zhang2019, salma2019}. Nevertheless, there are still open security gaps that require appropriate controls to mitigate them.
The challenge is that the currently proposed systems do not provide a complete security solution that tackles all IoT security and privacy requirements. For instance, most of the proposed methodologies target one or two security requirements, e.g., confidentiality and authentication \cite{Hamad2020}. An efficient and reliable IoT data sharing requires an all-inclusive security solution for securing the data while limiting the interference that might occur if integrating several independent techniques to provide the required services. To the best of our knowledge, none of the proposed research methodologies or industry systems  contribute a security attack free solution that provides conditional anonymous authentication and fine-grained access control techniques to be used by the resource-constrained IoT devices and infrastructures. 

There are a number of challenges confronting the security of IoT infrastructure, including but not limited to, scalable security, denial of use of service or upload of data, and interoperability. Scalability is one of the indispensable requirement in the IoT infrastructures. Such a requirement can be met by delegating the expensive cryptographic computations in a secured manner to a cloudlet, edge, or cloud \cite{Yang2016, Nguyen2018}. There is thus a dire need for investigating intelligent ways to use edge computing with IoT and the cloud to address the current security challenges of IoT systems. Moreover, IoT security research studies should consider using cryptographic methodologies with limited communication overhead, such as constant size Attribute Base Encryption techniques \cite {Belguith, Zhiyong2018}. 

Non-repudiation is another essential requirement for IoT infrastructure, specifically for systems that include users' interaction. Non-repudiation should be imposed to prevent users from denying either the use of the service or previous data upload. Unfortunately, non-repudiation is generally not considered in most of the current implementations owing to privacy concerns \cite{Li2015}. Several methodologies can maintain both users' and devices' privacy while implementing non-repudiations such as conditional anonymity. Group signature \cite {10.1007/3-540-46416-6_22, Khader2007} is one of the techniques that can provide conditional anonymity. However, such techniques require extensive research to concur with the limited resources challenge within IoT infrastructures.
Besides, {\em interoperability} is a vital IoT infrastructure requirement due to the heterogeneous nature of IoT devices. Considerable efforts and collaborations from governmental and non-governmental entities are required to create IoT interoperability standards and backward compatibility. These standards should also be integrated with privacy controls to guarantee the preservation of users' privacy. 

Trust is also an indispensable issue in the IoT environment since the majority of the existing security mechanisms do not cater for the subjective belief among the heterogeneous IoT objects especially in the presence of internal malicious adversaries that intent to disrupt the reliability of a network \cite{6468021}. Chaker et al. \cite{7797493} delineates on trust as the degree of a subjective belief of an entity (trustor) over the other (trustee) in a specified context. IoT interplays between the paradigms of security and trust, i.e., if we regard security mechanisms in terms of barriers, locks, and accesses, then trust is a worry of when, where, and why to put these barriers, locks, and accesses in an IoT ecosystem to deal with the degree of collaboration and integration between the IoT objects \cite{harwood2012logic}. Over the past decade, trust-based security mechanisms have emerged for enhancing the overall security of IoT, wherein trust and reputation models have been utilized to improve the collaboration and for selecting the trustworthy service provider based on quality-of-service (QoS), especially in service oriented architecture-based IoT~\cite{tcafe2019,6940301}. 

The importance of trust management has been 
recently investigated across numerous sort of networks, i.e., mobile ad hoc networks \cite{8319950}, peer-to-peer networks \cite{7407618}, social networks \cite{8737469}, and as-of-late for vehicular ad hoc networks within the context of the promising paradigm of Internet-of-Vehicles \cite{8730675}. Evaluating trust becomes indispensable in the case of a highly dynamic and distributed network since pervasive infrastructure cannot be guaranteed at all the times in such scenarios which is imperative for public-key based cryptographic techniques. Nevertheless, computing trust has its own inherent challenges, i.e., selection of dynamic trust attributes in accordance with a given application's context, assigning of optimal weights to such attributes for trust aggregation purposes, opting between the event-driven, time-driven, or hybrid approaches for trust updates, and selecting an appropriate trustworthiness threshold for segregating between malicious and non-malicious nodes. 

In essence, security, privacy, and trust go hand-in-hand for designing a resilient IoT network that could meet the stringent application requirements in realizing the formulation of highly secured digitized societies.

\subsection{Service Computing and IoT}
Initiated around the similar time as the Internet of Things, service computing (or service-oriented computing) has been
established as an important paradigm to change the way of design, delivery, and consumption of software applications~\cite{ShengQVSBX14,BouguettayaSHSD17}. Service computing relies on service-oriented architecture (SOA) and aims to organize software applications and infrastructures into a set of interacting services, which are then used as fundamental elements to support low-cost and efficient development of distributed applications. 

Technologies on service computing (e.g., RESTful services and service composition methods) can help address several fundamental challenges presented by IoT including communication and management of IoT objects. However, marrying service computing and IoT presents challenges due to their technical constraints and unique characteristics~\cite{BouguettayaSHSD17}. On the one hand, IoT objects may be resource-constrained and the traditional service computing standards and techniques (e.g., SOAP, WSDL, BPEL) might be too heavy to be applicable in IoT. On the other hand, existing service composition models cannot be directly used for IoT interoperation, due to their architectural differences. 
More specifically,  traditional service composition models are mostly single-typed and single-layered (i.e., services), while IoT components are heterogeneous, multi-layered that  include not only services, but also IoT devices and other components.   

One important research direction centers on {\em IoT services discovery}, aiming to be able to find the right IoT services at the right time and the right location. There are two possible techniques. The first technique is {\em semantic annotation} for IoT service descriptions and their associated sensory data. Some typical efforts in this direction include the OpenIoT project\footnote{www.openiot.eu}, which exploits a semantic sensor network (SSN) ontology from W3C, and the Hydra project\footnote{www.hydramiddleware.eu}, which adopts OWL (an ontology for Semantic Web) and SAWSDL (a semantic annotation of WSDL). However, it is challenging to reach an agreement on a single ontological standard for describing IoT services, given the diversity and rapid IoT technological advances. 
The second technical direction is to use the textual descriptions associated with IoT devices to locate IoT services. Some typical efforts in this direction include MAX~\cite{YapSM08} and Microsearch~\cite{TanSWL10}. One research challenge in this direction is the {\em natural order ranking} of IoT contents. Natural order ranking sorts contents by their intrinsic characteristics, rather than their relevance to a given query, thereby being able to deliver the most relevant results. One well-known example of natural order ranking is PageRank \cite{Page1999ThePC}, which orders Web pages based on their importance via {\em link analysis}. Given the size of IoT (50 to 100 times bigger than the current Internet), one promising direction is to develop a new natural order ranking mechanism for the IoT contents in order to provide an effective and efficient IoT service discovery~\cite{TranSBYZD19}. 

\subsection{Social IoT}
Recently, there have been quite a number of independent research activities to bring the next evolutionary step of the IoT paradigm by moving from smart objects to {\em socially aware objects}. This refers to creating a new generation of IoT objects that manifests themselves and have the capability to socialize with the surrounding peers mimicking human beings for the sake of, but not limited to, discovering new services, exchanging experience, and benefiting from each other capabilities.
This new paradigm is referred to as the Social Internet of Things (SIoT),  which is a new perspective that allows objects to establish their own social networks and navigate through the social network structure of the friend objects, allowing discovering other objects and their services. 

Unlike the current process in IoT where search engines are employed to find services in a centralized way, SIoT can foster resource availability and make services discovery more easily in a distributed manner 
\cite{atzori2012social,nitti2015friendship,ortiz2014cluster}.   
%
%
This paradigm also aims to provide reliable and trustworthy networking solutions by utilizing the social network structure. 
Based on the social structure established among IoT objects, objects can inquire local neighbourhood for other objects to assess the reputation of these objects and  establish a level of trustworthiness. Additionally, SIoT enables objects to start new acquaintance where they can exchange information and experience.


SIoT is not a spur of the moment. There were earlier attempts to involve devices in the social loop. Back to 2001, Holmquist et al. \cite{holmquist2001smart} established temporary relationships between wireless sensors. In the work of \cite{Bleecker}, the authors discussed the idea of how objects can blog. Moreover, kranz et al. \cite{kranz2010things} enabled objects to share content using a social network framework Twitter. Guinard et al. \cite{guinard2010sharing} utilized the human social network as a framework for owners to share the services of these devices with their friends. 

Previous attempts differ from the intended perspective of the current vision of SIoT. The current perspective refers to a new generation of IoT objects that have capability to form their own social network of friends without relying on the online human social networks. Several research activities have been conducted to realize this paradigm. In \cite{atzori2012social}, Atzori et al. introduced this new paradigm and discussed the idea of integrating social networks concepts into the Internet of Things (IoT) for the purpose of addressing the related issues of service discovery and composition. They proposed a conceptual platform on how to enable IoT objects to create relationships among each other. They also identified policies of how to establish relationships between objects and how to manage these relations. Girau et al. \cite{girau2017lysis} implemented an experimental SIoT platform. They evaluated the current implementations of IoT platforms and pointed out the major characteristics that can be reused in this experimental SIoT platform. It includes several functionalities that can enable the smart objects to register into the platform as a first step. Then, the system manages the creation of the new relationships. Using this system, smart objects are capable to create groups of members with similar characteristics. That leads to form a social network among each other by establishing social relationships autonomously with respect to the rules set by the owners. On the same research line, several studies have focused on proposing architectures \cite{ortiz2014cluster,voutyras2014architecture}. 

Relationships 
exist among smart objects. Objects can start establishing these relationships for several reasons such as when these objects come close to each other and satisfy relationships’ rules specified by their owners \cite{atzori2012social}. Atzori et al. \cite{atzori2012social} proposed five types of relationships. Some of these relationships are dynamic and they are established when smart objects come in contact at the same place and the same time periodically for cooperation to achieve a common goal. Other relationships are static and they are created once objects join the network. In addition, Roopa et al. \cite{roopa2019social} suggested extra relationships that can be established among objects.

Along with the previous research aspects, SIoT paradigm has gone through intensive research. Several SIoT areas such as service discovery \cite{bouyakoub2017smart,butt2018social,pham2015cloud,ruggeri2017framework},
network navigability \cite{militano2016enhancing,nitti2014network,nitti2016searching}, 
and trustworthiness management  \cite{nitti2012subjective,nitti2013trustworthiness,truong2017toward,xiao2015guarantor}  
have been studied in the literature. Furthermore, a recent work has considered how the SIoT resulted network would evolve since the SIoT network is dynamic where it can grow and change quickly over time where objects (nodes) and their relationships (links) appear or disappear \cite{aljubairy2020siotpredict}. However, the SIoT paradigm is still in an early stage, and there are many aspects that need to be investigated. Most importantly, the perspective of the SIoT paradigm needs to be thoroughly unified. In the future, IoT will be integrated more into daily life things and will have an interesting role to make decisions for humans. The couch in the living room could be able to sense the body temperature of the owner and based on this the room temperature gets adjusted accordingly. In another scenario, a smart medicine cabinet could monitor the consumption level of medicine and whenever the amount becomes low, this cabinet could ask the smart toilet to perform chemical analysis and report to the smart home in order to arrange a doctor visit or a refill from the pharmacy \cite{gershenfeld2000things}.

\subsection{IoT Recommendation}
With the exponential growth of data in the IoT environment, searching, accessing and connecting IoT devices are more difficult than ever. Therefore, a more desirable paradigm is proactively discovering suitable IoT devices rather than searching for one. In this new paradigm, instead of letting the user painstakingly searching for desirable devices to meet their needs, the automatic IoT system can suggest and deliver relevant resources to the user, matching with her history preferences. This IoT recommendation approach is an important research topic for the future applications of IoT, and we refer to it as the {\em thing-of-interest} (TOI) recommendation \cite{lina-thing-2016}. Due to the characteristics of the IoT environment, TOI has its own unique challenges and here we discuss three main challenges that TOI approaches have to overcome.

First, unlike common Internet resources such as document and images, IoT resources are inherently unreliable, ad-hoc, and not in uniform format \cite{lina-corre-2017}. We need reliable, trustworthy methods to be able to use these ephemeral and unorganized data. Hence, it is critical to understand the underlying relationships between IoT devices, to identify and group them together, and to aggregate data and reduce the unreliable nature of their data. Therefore, TOI services have to be dynamic and contextual-aware of their environment to keep track and quantify their IoT devices data sources. Second, 
    as sensors from IoT devices collect signal from surrounded environments, including personal human activities, privacy and security are of great concern when designing a TOI recommendation model. This challenge requires us to have new architectures and evaluation measurements regarding the performance of a TOI recommendation system, where the focuses are not only on the accuracy, but also the safety, security, and privacy of the involved entities. 
    Third, 
    the IoT environment is a distributed environment, while most of the recommendation approaches are run on a centralized server. This centralization nature does not fit well with TOI approaches, due to high demand traffic and aggregated data from cluster of IoT devices. This challenge requires new solution for TOI recommendation, and the recent trend is to deploy recommendation models on edge-devices such as mobile or portable IoT devices \cite{marjan-edge-2018}.

Given these challenges, TOI recommendation systems have a significant deviation from the normal recommendation approaches, and we must carefully address them. Furthermore, we 
envision new promising directions for future research in this area. Firstly, applying deep learning techniques to build TOI models are increasingly necessary. Deep learning methods can draw out complex patterns and behaviors of IoT device's signals, thus are very helpful for context-aware TOI recommendation systems. The second promising direction is the interpretability of TOI recommendation \cite{shuai-deepsurvey-2019}. By achieving explainable reasons for the recommendation, the IoT system can convince its users for better adaptability, and help the users learn more insights from the decision rationale behind the recommendation. Another promising future direction is combining with IoT searching to have a more powerful recommendation system. By having both proactive and post-active approaches in a recommendation system, users can have better experience when looking for TOI. This combining approach is also an effective method to 
overcome the cold-start issue that has to be faced by the most recommendation systems.

\subsection{Edge Computing and IoT}

Over the past decade, 
an unprecedented increase in the deployment of IoT devices coupled with the 
demanding 
of
real-time computing power and low-latency requested by the state-of-the art applications continues to drive the case for edge-computing systems. Such applications include, but are not limited to, smart cities (with autonomous driving being its integral constituent), healthcare, augmented reality, robotics, and artificial intelligence.

Edge computing is primarily a part of the distributed computing topology which has an intent to bring both computation and storage near to the devices.
This is quite beneficial for applications requiring stringent latency requirements. For instance, in case of safety-critical vehicular applications, i.e., forward collision warnings, lane changing assistance, emergency vehicular assistance, and blind intersection warnings, a maximum tolerable threshold of 3-10 milliseconds is indispensable for mitigating performance-related issues \cite{Adnan-MDPI,B.Ji-IEEECSM}. This is also economical and resource efficient considering the fact that most of the data is processed itself at the edge and only a handful of data is sent back to the centralized cloud, thereby reducing bandwidth requirements.  

Nevertheless, a number of edge-based IoT applications are a source of momentous amount of data, e.g., as per an estimate of Automotive Edge Computing Consortium \cite{AECC}, connected vehicles are anticipated to generate an approximately 5 TB of data for every hour of their driving with a large chunk of the same transpiring from the video cameras primarily employed for computer vision purposes in order to facilitate vehicles to gain a perception of the world around them. With the advent of 5G and beyond 5G wireless communication technologies, data volumes continue to grow since more and more sophisticated edge-based IoT devices are being seamlessly integrated in the network. In addition to connected vehicles, numerous sensors and roadside infrastructure within the context of the paradigm of smart cities, handheld devices (cellular and other computing devices), home automation devices installed in smart homes, and intelligent bots operating on the factory floor all constitute edge-based IoT devices. In order to intelligently manage such big data, developing an edge-centric data management strategy with highly specialized analytics capabilities is hence of the essence in order to glean insights in real-time with fairly limited computing power. By effective decentralized decision making, edge analytics is capable of identifying a cause well before its respective effect has actually materialized. 

This unprecedented growth in the number of edge-based IoT end points also results in an increase in the attack surface, i.e., an aggregate of a system’s end points which an attacker could leverage for his malicious gains. Therefore, security is of the most pressing concerns for the edge since IoT devices which connect to the public Internet largely results in compromising the security protocols. This all boils down to the current state of the edge computing since full stack solutions encompassing sensors, software, and secure elements are almost non-existent. IoT networks at the edge further rely on the low-power wide-area network (LPWAN) protocols which themselves employ simple cryptographic techniques and are prone to attacks especially in case the encryption keys have been compromised. Moreover, VPNs are also subject to man-in-the-middle attacks. Nevertheless, implementing an end-to-end encryption and creating mechanisms for securing edge-based IoT devices via embedding security features within them (and in the edge data centers) 
would 
facilitate 
a resilient expansive network.  

\subsection{Conversational IoT}
The most natural way for human beings to interact is through words.
Advanced technology combined with extensive research over the past few years has made it possible for the humans to communicate with the machines using natural language, thus giving rise to the field of \textit{Conversational AI}. It refers to the use of either text-based or voice-based applications that enable machines to stimulate human conversations and create a personalized experience for the users. These conversational agents can be envisaged as a natural interface for the IoT devices as it hides all the complex applications, services and hardware such as sensors and actuators, presenting a daunting challenge of gaining technical knowledge to interact with the various components. 

The convergence of IoT and Conversational AI is regarded as successful as we have seen many 
applications already making their way to people's customized smart spaces such as smart offices, smart homes, and smart vehicles. The first in the line of transformation of a regular home into a smart IoT home is `Google Home', which is flexible to work with and provides a centralized solution to control 
compatible smart home devices. 
Another state-of-the-art device is Amazon's Echo which provides more improved features than Google Home. Echo can guard a home in owner's absence by listening to surroundings for unusual noises or alarms. 
A `Home and Away' feature can be set up to trigger specific actions.
It helps shop from Amazon and notifies the owner when the parcel arrives. In a multi-user environment, each user can register their account using voice activation.  Though these devices and alike (Alibaba Group's Tmall Genie etc.) overcome interoperability issues to make the life an effortless, seamless experience, 
they suffer from a number of limitations due to which a huge performance gap is easily observable on managing the smart spaces as a whole. 

These limitations can be considered as potential research challenges which include, but are not limited to: i) \textit{Self-Disclosure in a Multi-User Environment:} the increase in the number of interactions between the system and end-users would result in increased disclosure about user's activities and personal information. This disclosure of information helps the system in understanding the user and thus aids in providing a more personalized experience. However, in multi-user scenario this disclosure of personal information may poses high risks pertaining to one's privacy and thus, requires a model where multiple users can co-exist without having to worry about security or data breach; ii) \textit{Lack of Complexity and Completeness:} the available IoT conversational agents work on simple commands like ``turn on the TV" or ``what is the temperature of the room?". However, these systems struggle with rules or complex sentences such as ``turn off the heater when the room is warm" or ``turn on the TV when Prison Break is on", unless the user decomposes them into separate simple sentences \cite{lago2020conversational}. Thus, extensive research is required to make the systems tackle incomplete or complex sentences without having to decompose them to keep the conversation natural; 
iii) \emph{Inability to Reason:}
Commonsense reasoning is considered as a key factor to the success of many natural language processing tasks specifically in question answering and conversation dialog \cite{lin2017reasoning,wang2017conditional}. The machine should be able to provide rationale answers to questions like ``Why is it so cold today?" in order to establish effective interactions. 
Unfortunately, current technologies are still far from realizing this capability and more research efforts are needed. 
iv) \textit{Lack of Conversational Context:}
the more natural and interactive way of having a conversation is by incorporating historical context into the conversation. Consider an example \textit{\{User: Who is the most dialed number in my call record?, Agent: Emma Collins, User: Could you please set her as my emergency contact?\}}. The agent needs to maintain the record of turn 1 in order to decipher `her' in Question 2. Most of the IoT conversational agents are single-turn agents where they do not keep track of the previous conversation, and thus provide inaccurate answers \cite{DBLP:conf/emnlp/ChoiHIYYCLZ18,reddy2019coqa}. Thus, designing agents that keep track of the previous turns is an important research direction. 


\subsection{Summarization in IoT}

With advances in the Internet of Things, the proliferation of data generated from 
sensors and the growth of Internet users have created a pressing need for compressing the data over the Internet. 
Textual
data is one of such data. From natural language data processing perspective, summarization is an effective technique for data aggregation that can generate a short and concise summary from one or one set of texts. In the IoT era, documents are located in a distributed way, raising the research of multi-document summarization \cite{corr/abs-2011-04843}.

Towards this end, combining IoT and summarization technology is worthy to be explored. More specifically, data collected from IoT networks 
are processed by summarization techniques. Eventually, condensed semantical features are formed, with which the downstream tasks will be facilitated dramatically. By doing so, it can help the IoT users save huge amount of time, since the users are able to quickly acquire target information they need without reading tedious documents \cite{ji2014big}. Moreover, data summarization is capable of reducing the energy consumption in various IoT environments \cite{nithyakalyani2013data} and decrease the requirements of the application servers in storage, transmission and processing.

The combination of IoT and the summarization technology could have a wide range of applications. For instance, a request to some intelligent devices, such as Google Home, is given by IoT users to fetch highly condensed information. The devices would search around the Internet to gather the most relevant documents; later on, text summarization techniques will be performed to pick the key points out of tons of information to form the final concise answers. Not only text summarization techniques can be applied to many scenarios, video summarization techniques can also be combined with IoT in the seek of fast and efficient information processing \cite{muhammad2019cost}. For example, in intelligent security areas, the surveillance videos can be summarized by video summarization algorithms to extract the most informative and important features. These techniques are able to be applied in smart cities as well, where traffic videos could be obtained and video summarization algorithms could play an important role. 

Besides text and video data processing, in recent years, with the tremendous successes gained by data-driven approaches, multi-modal data processing attracts increasing attentions. These data come from texts, audios and videos, which provide a more comprehensive view. Multi-modal data processing enables models to fuse data from different sensors and sources, but it will inevitably incur exponentially increasing data to be processed. Under this circumstance, summarization algorithms to process multi-modal data can be adopted to fuse information semantically. Despite the advantages of the combination of summarization techniques and IoT,
it is still a new area, with very few existing works. 
Deep neural networks with conventional text and video processing techniques can be investigated, since deep neural models have strong non-linear mapping abilities and traditional approaches contains many prior knowledge, which would facilitate the model optimization process. 
We foresee that 
summarization on IoT will be one of the next intensely researched topics.

\section{Conclusion}
The Internet of Things (IoT) has been an extremely active area of research and development for more than two decades. Although a wealth of exciting activities including standardization, commercial developments and research have been conducted, many challenges still remain open due to the large scale and diversity of IoT devices, the openness of the IoT environment, and the security and privacy concerns. In this paper, we identify 10 key research topics on IoT and hope to stimulate further research in this vibrant area.

\section*{Acknowledgment}
Some viewpoints of this invited paper are from previous papers by the authors, in particular~\cite{Hamad2020,Zeadally-EH,csur/TranSBY17,BouguettayaSHSD17,QinSFDWV16}. 
Quan Z. Sheng's work has been partially supported by Australian Research Council (ARC) Discovery Project Grant DP200102298, LIEF Project Grant  LE180100158, 
and Future Fellowship Grant FT140101247.

\bibliographystyle{IEEEtran}
\bibliography{michaelSheng}

\end{document}